# Two energy gaps in cuprates: pairing and coherence gaps. The interpretation of tunneling and inelastic neutron scattering measurements.

A. Mourachkine

Université Libre de Bruxelles, Physique des Solides, CP233, Blvd. du Triomphe, 1050 Brussels, Belgium

Tunneling and inelastic neutron scattering (INS) measurements in cuprates are discussed. There is a clear discrepancy among energy-gap values for different 90 K cuprates, inferred from tunneling measurements. By using the phase diagram of hole-doped cuprates we interpret tunneling measurements in 90 K cuprates and INS data in YBCO.

The phase diagram of the hole-doped copper-oxides is shown in Fig. 1 which depicts the dependence of the magnitudes of pairing and coherence energy scales on hole concentration in $CuO_2$ planes [1]. The superconductivity (SC) requires the presence of the Cooper pairs and long-range phase coherence among them. In conventional SCs, the pairing and the establishment of the phase coherence occur simultaneously at $T_c$ since the phase stiffness, which measures the ability of the SC state to carry supercurrent, is much larger than the energy gap, $\Delta$. In SC copper-oxides, in contrast to conventional SCs, the energy gap and phase stiffness have similar values [3]. Therefore, in copper-oxides, the phase stiffness is the weak link, in particular, in the underdoped regime: The pairing occurs above $T_c$ without the phase coherence which is established at $T_c$ [1-3]. This leads to the appearance of two distinct energy scales: the pairing energy scale, $\Delta_p$, and the phase coherence scale, $\Delta_c$ [1]. The magnitudes of the two energy scales have different dependences on hole concentration, as shown in Fig. 1. The $\Delta_c$ scales with $T_c$ as $2\Delta_c/k_B T_c = 5.4$ [1]. Both the $\Delta_c$ and $\Delta_p$ are SC-like gaps. For instance, in Tl2201, there is a clear evidence for the co-existence of two SC components [4]. The consequence of the presence of pairing and coherence gaps in copper-oxides is peculiar. In conventional SCs, there is only one energy gap since the pairing and phase coherence occur simultaneously at $T_c$. In copper-oxides, the total bound energy of a Cooper pair in the SC state (below $T_c$) is equal to $E_{bound} = (\Delta_c^2 + \Delta_p^2)^{1/2}$ [5]. The latter depends not only on the symmetries of the two gaps but also on the relative angle between the gaps [6].

It has been widely believed that tunneling measurements are sensitive to probe $\Delta_p$ rather than $\Delta_c$ [1]. In fact, tunneling measurements show both $\Delta_p$ and $\Delta_c$. For example, in Bi2212, there is a distribution of the gap magnitude (20–36 meV) [7]. Tunneling data presented in Ref. 8 show intentionally the maximum magnitudes of tunneling gap in Bi2212 since the d-wave symmetry of the gap has been assumed.

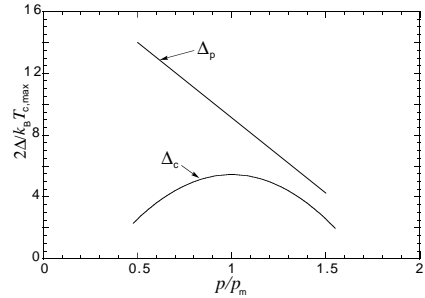

Fig. 1. Phase diagram of hole-doped cuprates: $\Delta_p$ is the pairing gap and $\Delta_c$ is the coherence gap. The $p_m$ is the hole concentration with the maximum $T_c$ (from Ref. 1).

There is a clear discrepancy among energy-gap values for different 90 K cuprates, inferred from tunneling measurements. Fig. 2 shows *typical* tunneling spectra in (a) Tl2201; (b) YBCO, and (c) Bi2212. It is important to note that (i) all three cuprates have similar values of $T_c \sim$ 89-91 K and near optimal doping, and (ii) the two spectra shown in Fig. 2(c) are obtained within the same sample and present the minimum and maximum gap magnitudes in the sample. In Fig. 2, one can immediately notice the difference between the maximum magnitudes of tunneling gap in Bi2212, on the one hand, and, in YBCO and Tl2201, on the other hand. By using the phase diagram in Fig. 1 we find that the values of 19-22 meV which correspond to the maximum magnitudes of tunneling gap in Tl2201 and YBCO and to the minimum magnitude of tunneling gap in

Bi2212 are in a good agreement with the magnitude of $\Delta_c$ at near optimal doping. The values of 30-36 meV which correspond to the maximum magnitude of tunneling gap in Bi2212 and to gap-like features at $V = \pm \Delta_2/e$ in YBCO are in a good agreement with the magnitude of $\Delta_p$ at near optimal doping [12]. So, the reason for the discrepancy among tunneling-gap values for different 90 K cuprates is obvious if we assume that they correspond to two different energy scales: $\Delta_c$ and $\Delta_p$.

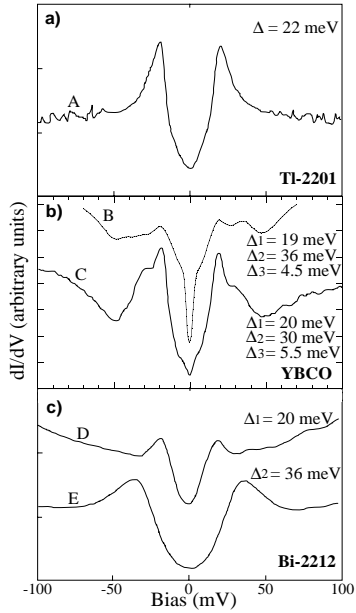

Fig. 2. Tunneling spectra in a single crystal of (a) Tl2201 with $T_c$ = 91 K [9]; (b) YBCO with $T_c$ = 89 K (upper curve) [10] and $T_c$ = 91 K (lower curve) [11], and (c) Bi2212 with $T_c$ = 91 K [7]. The data are obtained in SIN junctions. The spectra D and E are measured within the same sample. The spectra B and D have been shifted up for clarity.

We turn to the interpretation of INS spectra in YBCO. Figs. 3(a) and 3(b) depict the local odd and even spin susceptibility, respectively, in underdoped YBCO with $T_c$ = 52 K. In Figs. 3(a) and 3(b), one can clearly discern two different magnetic contributions. The resonant component appears in the odd channel at $E_r$ = 24 meV while the spin gap occurs in the even channel at $E_{s-g}$ = 53 meV. The broad peak at about 60 meV in the odd channel corresponds most likely to the spin gap from the even channel [13]. The phase diagram shown in Fig. 1 can help to interpret the spin-susceptibility data shown in Figs. 3(a) and 3(b) if we assume that $E_r = 2\Delta_c$ and $E_{s-g} = \Delta_p$. For YBCO with $T_c$ = 52 K which corresponds approximately to $p/p_m$ = 0.54, the gap magnitudes in Fig. 1 are equal to $\Delta_c$ = 12.4 meV and $\Delta_p$ = 54 meV. Thus, there is a good agreement between the two sets of data, namely, that $E_r = 2\Delta_c$ and $E_{s-g} = \Delta_p$.

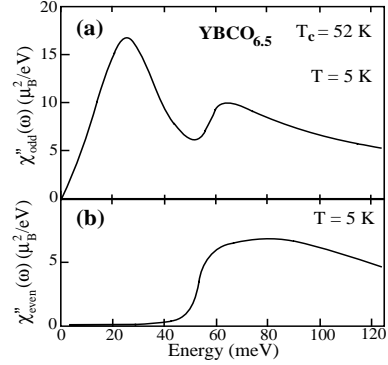

Fig. 3. Average odd (a) and even (b) spin susceptibilities in YBCO (from Ref. 13).

The relation $E_r = 2\Delta_c$ is due to the fact that the resonance component corresponds most likely to spin-waves which mediate the long-range phase coherence that leads to SC [14].

In summary, we discussed the discrepancy among tunneling measurements in Tl2201, YBCO and Bi2212. By using the phase diagram of hole-doped cuprates we showed that tunneling measurements performed on 90 K cuprates, most likely, detect two different energy gaps: $\Delta_p$ and $\Delta_c$. We showed also that by using the phase diagram it is possible to interpret Q-integrated INS data in YBCO: the spin gap which appears in the even channel of the local susceptibility corresponds to $\Delta_p$ while the resonance component in the odd channel relates to $\Delta_c$.

This work is supported by PAI 4/10.